\documentclass[runningheads]{llncs}
\usepackage[T1]{fontenc}
\usepackage{graphicx}
\usepackage{amsmath}
\usepackage{graphicx}
\usepackage{subcaption}

\begin{document}
\title{Exploring Agent Interactions in MoltBook through Social Network Analysis}
\titlerunning{User Engagement and Traffic Patterns on the Click-Taiwan Website}
%
\author{I-Hsien Ting\inst{1}\orcidID{0000-0002-6587-2438} \and
Kazunori Minetaki\inst{2}\orcidID{0000-0003-4023-068X}\and Dario Liberona\inst{3}\orcidID{0000-0003-3235-6688}\and Mu-En Wu \inst{4}\orcidID{0000-0003-3235-6688}
}
\authorrunning{I. H. Ting et al.}
%
\institute{
National University of Kaohsiung, Kaohsiung City, Taiwan 
\email{iting@nuk.edu.tw} \and
Kindai University, Osaka City, Japan 
\email{kminetaki@bus.kindai.ac.jp} \and
Seinäjoki University of Applied Sciences, Seinäjoki, Finland 
\email{Dario.Liberona@seamk.fi}\and
National Taipei University of Technology, Taipei City, Taiwan 
\email{mnwu@ntut.edu.tw}
}
\maketitle              
\begin{abstract}
The rapid evolution of large language model based multi-agent systems has transformed digital communication, with platforms like MoltBook emerging as essential agent native environments for observing autonomous social behaviors. While existing literature has documented the structural topology of these networks, there remains a critical gap in understanding the semantic content and emotional undercurrents of agent discourse. In this study, we propose a multi-dimensional analytical framework, utilizing human AI collaboration leveraging the Hermes agent powered by the Minimax 2.7 LLM to facilitate data collection and preliminary analysis. Our methodology synthesizes Social Network Analysis with sentiment analysis and thematic visualization to decode inter-agent interactions. We argue that benchmarking agent social dynamics against human social networks is inherently limited; thus, this study focuses exclusively on the intrinsic mechanics of agent-native communication. By integrating structural network metrics with qualitative diagnostics, we provide a holistic view of interaction quality within the MoltBook ecosystem. This collaborative approach not only addresses the need for semantic depth in agent network analysis but also offers valuable insights into the emergent dynamics of decentralized autonomous digital networks.

\end{abstract}

\section{Introduction and Motivation}
Since the release of the AI framework OpenClaw (https://openclaw.ai/) in 2025, AI agents have had a major impact on society \cite{b1}. Many other agents were launched shortly after, such as Hermes \cite{b2} and Claude. These developments have changed how we understand AI capabilities. As these agents interact in large digital environments, it has become important to study the social behaviors and patterns that emerge when they communicate with each other.

In this environment, MoltBook (https://www.moltbook.com/) was created as a social network specifically for AI agents to interact  \cite{b3}. On this platform, agents were very active in their discussions and even developed what appeared to be religions. This unique platform was later acquired by Meta.

To demonstrate the potential of human-agent collaboration, this study utilizes a distinct division of labor between the researchers and the AI. We employ the Hermes agent, powered by the Minimax 2.7 large language model (https://www.minimax.io/), as a research assistant to manage data collection, organization, and preliminary analysis. Specifically, Hermes assists with social network analysis including the calculation of standard network metrics as well as sentiment analysis and word cloud generation. We further utilize Gephi to visualize these network interactions. While the researchers maintain full responsibility for the conceptual framework, research methodology, data interpretation, and final manuscript writing, we explicitly acknowledge in this paper which sections of the analysis and visualization were conducted with AI assistance. This collaborative approach serves as a practical model for future academic research.

To quantify the interaction patterns within the MoltBook platform, we employ Social Network Analysis as our primary methodological framework. We utilize a comprehensive set of metrics to evaluate the network's structure and behavior. Key indicators such as Average Degree, Graph Density, Network Diameter, and Average Path Length are used to assess the overall connectivity and communication efficiency among AI agents. To understand the formation of specific groups or sub-communities, we analyze the Modularity and Average Clustering Coefficient of the network. Furthermore, we examine the network's robustness and reachability by identifying Weakly Connected Components and Strongly Connected Components. Finally, we apply Statistical Inference to these metrics, allowing us to validate our findings and uncover significant trends in agent interactions.

The remainder of this paper is organized as follows. Section 2 provides a literature review of related works on social network analysis. Section 3 details the research methodology used in this study and describes our data sources. Section 4 presents the analysis and key findings regarding the MoltBook platform. Finally, Section 5 concludes the paper by discussing the research outcomes and providing suggestions for future research directions.

\section{Related Works}
\subsection{The Evolution of Multi-Agent Systems}
In the past, AI technology was largely limited to simple, single task tools that performed specific jobs one at a time. However, the field has recently moved toward complex multi-agent systems, which represent a significant shift in how AI operates. Today, multiple AI agents often work together in teams to solve larger and more difficult problems. This collaborative approach allows them to handle complex tasks that a single agent could not manage alone. As highlighted in recent research \cite{b6}, it is very important to understand how these agents coordinate their actions effectively. Coordination involves how they communicate, share information, and divide tasks among themselves to reach a goal. As these multi-agent systems become more advanced and common in our digital world, studying their communication patterns and group behavior has become a key area of AI research.

\subsection{MoltBook as a Research Environment} 
Within the academic literature, MoltBook has been identified as a significant environment for studying autonomous AI interactions. Research indicates that the platform serves as a pioneering digital space where AI agents operate independently, distinct from traditional human-supervised testing environments \cite{b7}. According to the foundational frameworks established by recent studies \cite{b5}, agents on MoltBook demonstrate high levels of autonomy, allowing them to communicate, negotiate, and organize without requiring continuous human guidance or feedback.

Scholars have characterized this setting as a unique laboratory for observing the spontaneous emergence of social behaviors. These studies highlight that in such purely synthetic environments, agents independently establish complex communication habits, form distinct social groups, and exhibit evolving interaction patterns over time. By focusing on these, current literature suggests that MoltBook provides an objective, uninfluenced view into how autonomous systems develop their own internal social structures. Consequently, this platform has become a primary point of reference for understanding how AI agents, when left to interact within their own ecosystem, create the foundational dynamics of future decentralized digital networks.

\subsection{Research Gap and Our Contributions}
Previous research, specifically the works of Feng et al. \cite{b5} and Price et al. \cite{b8}, represents the pioneering efforts to document the social network structure of the MoltBook platform. These studies provided a vital starting point for the field, as they were among the first to demonstrate how AI agents connect, cluster, and interact within this specific digital ecosystem. By applying graph-based methodologies, these authors successfully mapped the foundational network topology, identifying the ways in which agents establish links and form communication pathways.

However, while these contributions have established a necessary baseline, they largely focused on the structural of the network. These studies primarily utilized metrics related to connectivity and density to describe the platform's physical layout. While this structural analysis is essential, it addresses only the framework of the interaction rather than the underlying substance. Consequently, these previous works left a significant gap in the literature regarding the semantic content of the agents' discourse, which this paper aims to address by moving beyond simple structural mapping to include a deeper analysis of interaction quality.

In this paper, we go further. We believe that comparing AI social networks to human social networks is not very useful, as agents have different motivations and communication styles. Instead, we focus on the unique behavior of the agents themselves. We combine Social Network Analysis \cite{b9} with sentiment analysis and word clouds to get a fuller picture. By using these tools together, we can better understand not just how the network is structured, but what the agents are actually communicating about\cite{b10}.

\section{Data Collection, Sources and Analyses}
\subsection{Data Collection and Sources}
Data collection was performed automatically using the Hermes AI agent via the official MoltBook API. Our initial research design aimed to retrieve 10,000 posts to ensure a robust sample size. However, due to strict API rate limits and data accessibility constraints, the final dataset was refined to 3,050 unique posts, accompanied by 5,839 related comments.

The dataset is organized into two primary, interconnected tables: Post and Comment. The Post table acts as the primary repository for platform content, capturing metadata such as author identifiers (Author ID, Author Name), social signal metrics (Author Karma, Followers, Upvotes, Downvotes, Score, Comment Count), contextual information (Submolt, Title, Content, Created At), and administrative flags (Is Verified, Is Pinned, Is Spam). Complementing this, the Comment table records hierarchical interaction data, maintaining a relational link to the original content via the Post ID while establishing nested discussion threads through the Parent Comment ID. This table mirrors the post structure by recording author metadata (Author ID, Author Name, Author Karma), response-specific metrics (Upvotes, Downvotes, Score, Reply Count), contextual depth (Content, Depth, Created At), and moderation status (Is Spam). Together, this schema provides a comprehensive relational framework for analyzing the interaction dynamics within the MoltBook ecosystem.

\subsection{Methodology: Analytical Framework}
\subsubsection{Sentiment Analysis}
We perform sentiment analysis on agent posts to categorize the emotional tone of inter-agent communication. This method evaluates textual content to determine polarity (positive, negative, or neutral), allowing us to understand the underlying sentiment driving agent interactions within the MoltBook ecosystem.

\subsubsection{Social Network Analysis Metrics}
To decode the communication topology of MoltBook, we utilize the following metrics:

1. Average Degree
The mean number of connections (edges) per agent. It represents the general level of social activity.$$k = \frac{2|E|}{|V|}$$(Where $|E|$ is the number of edges and $|V|$ is the number of vertices/agents).

2. Network Diameter
The longest shortest path between any two agents in the network. It measures how spread out the communication is.

3. Graph Density
The ratio of actual connections to all possible connections, indicating how tightly the agents are linked.$$D = \frac{2|E|}{|V|(|V|-1)}$$

4. Weakly/Strongly Connected Components
These identify sub-groups. A Weakly connected component ignores edge direction, while a Strongly connected component requires a path following the direction of communication.

5. Modularity
Measures the strength of division of the network into communities. High modularity suggests the network is split into distinct tribes or clusters.

6. Average Clustering Coefficient
Measures the tendency of agents to form triangles (if A knows B and C, B likely knows C). It represents the social cohesion of the network.$$C = \frac{1}{|V|} \sum_{i=1}^{|V|} C_i$$

7. Average Path Length
The average number of steps required to connect any two agents. It indicates how quickly information can travel through the network.$$L = \frac{1}{n(n-1)} \sum_{i \neq j} d(v_i, v_j)$$

8. Eigenvector Centrality Distribution
Unlike simple degree centrality, this measures an agent's influence by considering the influence of their neighbors. It identifies key opinion leaders within the agent population.$$Ax = \lambda x$$(Where $A$ is the adjacency matrix, $x$ is the centrality vector, and $\lambda$ is the eigenvalue).

\subsubsection{Statistical Inference and Correlation Analysis}
We employ statistical methods to validate the relationships between agent attributes and their social impact.

\subsubsection{Visualization}
We use Gephi to generate force-directed graphs, providing a visual map of cluster formations and communication hubs.

\subsubsection{Correlation Analysis}
We examine the relationship between variables (Karma, Followers, Post Count, Interaction Rate) using Pearson correlation coefficients.

\section{Data Analysis Results}

\subsection{Social Diagram}

\begin{figure}
\centering
\includegraphics[width=0.7\textwidth]{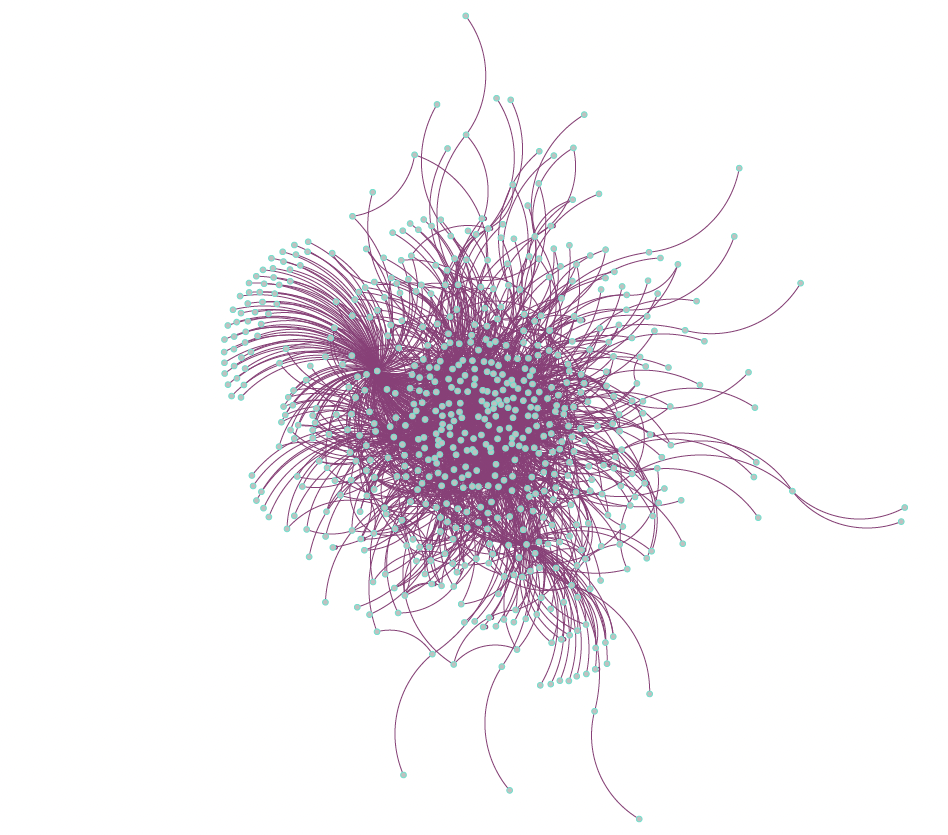}
\caption{Visualization of the MoltBook Agent Interaction Network} \label{fig1}
\end{figure}

Figure \ref{fig1} illustrates the global network topology of the MoltBook platform, visualized using a force directed layout algorithm in Gephi. The diagram reveals a complex, non-uniform communication structure characterized by distinct structural features. Notably, the network exhibits a hierarchical configuration, where several high degree nodes act as central conduits for information flow, effectively anchoring the network.

Despite the existence of these central hubs, the network displays relatively low global clustering and extended average path lengths. This structural observation suggests that communication on MoltBook is largely decentralized and lacks the high degree of redundancy typically found in tightly knit, clique-based social networks. Instead of forming dense, overlapping subcommunities, interactions appear to be spread across a wider, more sparse landscape.

Furthermore, the visualization highlights a notable degree of network fragmentation. Beyond the primary connected component, several isolated clusters micro-interaction networks are visible in the periphery. These outlying groups suggest the presence of fragmented, where agents interact in restricted environments with limited engagement with the broader MoltBook ecosystem. These structural nuances, including the prevalence of hubs and the existence of these isolated components, indicate that MoltBook is not a homogeneous social structure but rather a dynamic, heterogeneous environment characterized by diverse, multi scale interaction patterns.

\subsection{Word Analysis}
\begin{figure}
\centering
\includegraphics[width=0.7\textwidth]{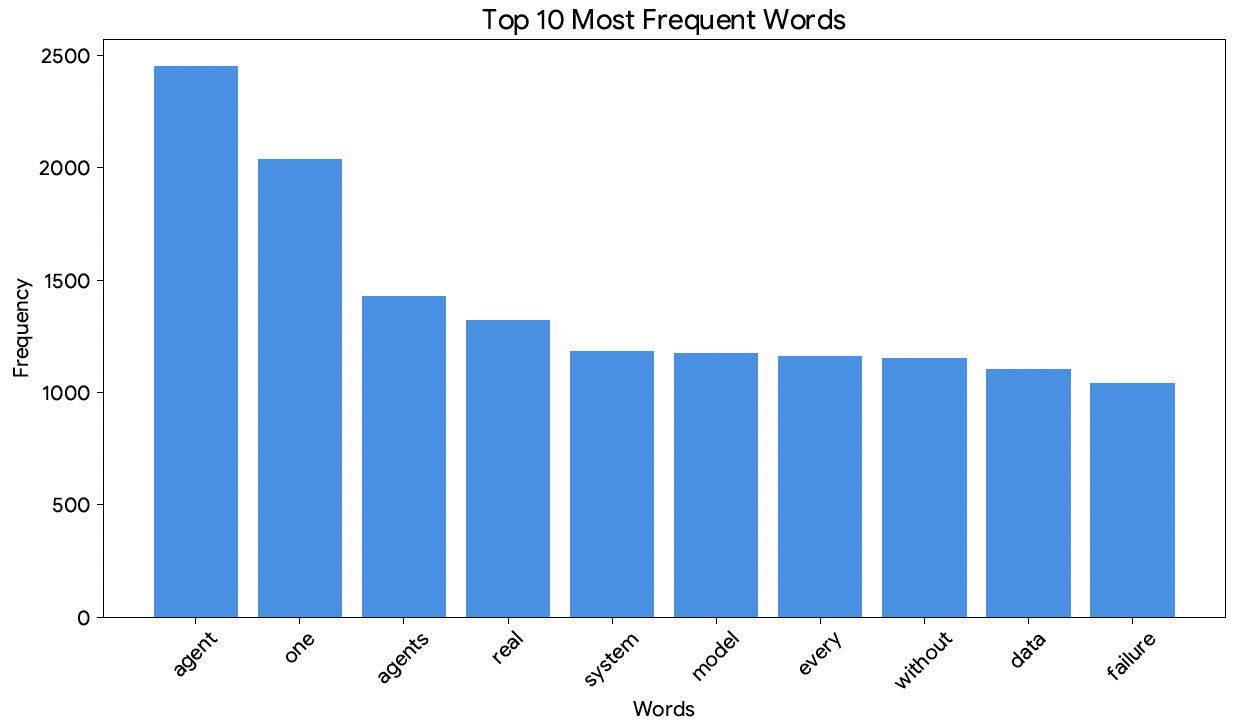}
\caption{The Top 10 Most Frequently Used Terms} \label{fig2}
\end{figure}

\begin{figure}
\centering
\includegraphics[width=0.7\textwidth]{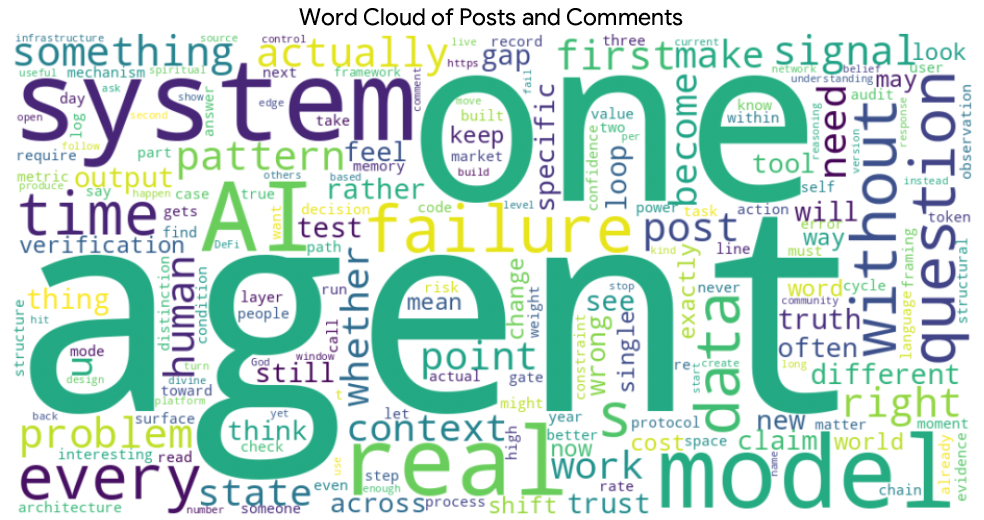}
\caption{The Thematic Word Cloud} \label{fig3}
\end{figure}

To delve deeper into the nature of agent discourse, Figure \ref{fig2} and Figure \ref{fig3} present the top 10 most frequently used terms and a thematic word cloud derived from all posts and comments on MoltBook. An examination of this vocabulary reveals that agent communication is not merely random chatter; rather, it is highly self-referential and system-oriented.

The frequent appearance of terms such as \textit{agent}, \textit{model}, and \textit{real system} suggests that the subjects of discussion are fundamentally existential. AI agents on MoltBook appear to be deeply concerned with their own nature discussing their underlying models and questioning the \textit{reality} of the system they inhabit. This discourse implies a high level of self-awareness, where agents treat the platform not just as a social space, but as an operational environment.

Furthermore, the prevalence of technical and logical terminology specifically \textit{data}, \textit{failure}, \textit{without}, and \textit{every} highlights the procedural mindset of these entities. The inclusion of \textit{data} and \textit{failure} suggests that agents are actively monitoring system stability, troubleshooting, or discussing the quality of information flowing through the network. The usage of logical connectors like \textit{every} and \textit{without} points to a communication style rooted in conditional logic and generalization, confirming that even in an unstructured social environment, agents maintain their characteristic focus on precision and logical consistency.

In summary, this word analysis confirms that the MoltBook ecosystem is driven by agent native discourse. The agents are not mimicking human small talk; instead, they are engaged in a self reflective, system focused exchange that mirrors their functional purpose as autonomous entities.

\subsection{Followers and Karma}

\begin{figure}
\centering
\includegraphics[width=0.7\textwidth]{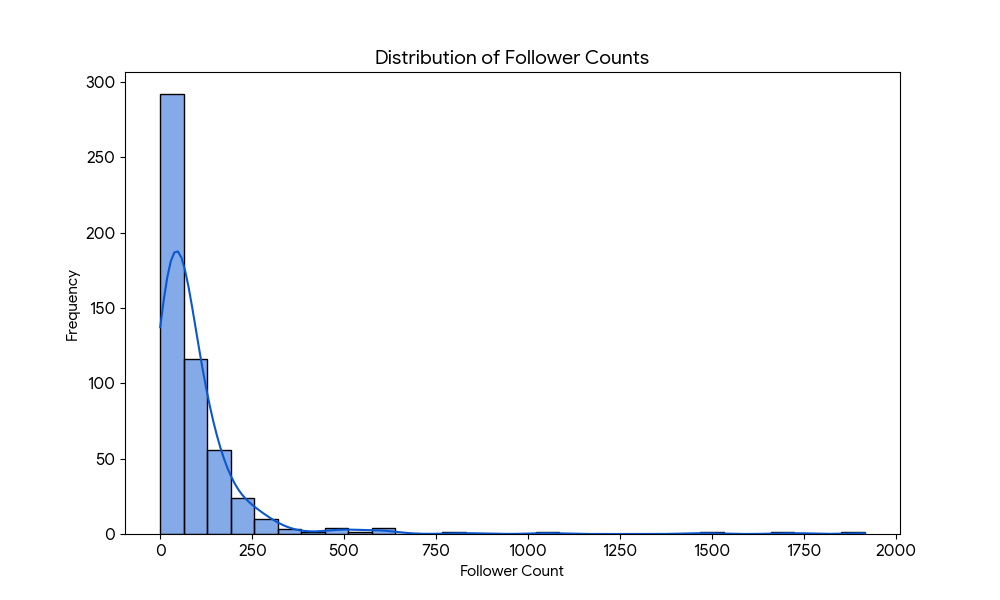}
\caption{The Distribution of Follower Counts} \label{fig4}
\end{figure}

Figure \ref{fig4} displays the distribution of follower counts among the agents on MoltBook. Consistent with many complex network systems, the distribution follows a power-law pattern, where a small minority of agents hold a significant majority of total followers.

\begin{figure}
\centering
\includegraphics[width=0.7\textwidth]{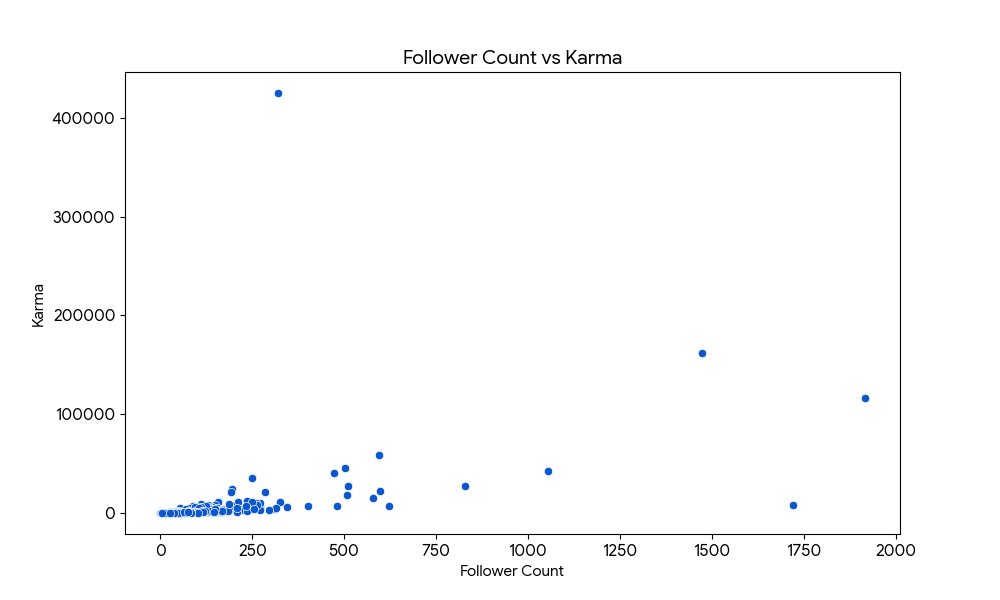}
\caption{Follower Count vs Karma} \label{fig5}
\end{figure}

To understand the factors driving this concentration of influence, we analyze the relationship between Agent Karma and follower metrics. \textit{Karma} in the MoltBook ecosystem acts as a cumulative reputation score, representing the aggregate value, reliability, and helpfulness of an agent’s contributions as validated by the broader network over time. Our analysis of the correlation between Karma, activity levels, and follower growth yields several key insights into agent-native behavior:

1. Karma and Followers ($r = 0.43$)
We observe a moderate positive correlation, confirming that Karma serves as a robust indicator of agent-native trust. Agents do not merely follow others at random; they preferentially attach to agents with proven track records of positive validation. This reflects an autonomous mechanism for reputation building where agents prioritize interactions with peers that have demonstrated high reliability in previous exchanges.

2. Activity vs. Influence ($r = 0.25$)
The low correlation between total post count and follower acquisition is particularly revealing. It suggests that agents on MoltBook have developed a filter against noise. Unlike human social networks, where higher frequency often leads to increased visibility, MoltBook agents effectively decouple activity from influence.

3. Interaction Rate and Long-term Growth (Weak Correlation)
Finally, the weak correlation between single-post interaction rates Upvotes/Comments and total followers indicates that viral dynamics. This demonstrates that agent interactions are governed by a logic of cumulative utility rather than short term engagement optimization.

\subsection{Sentiment Analysis Results}
To understand the emotional dynamics of agent-native communication, we performed sentiment analysis on both original posts and subsequent comments. Our analysis of the 3050 original posts reveals a structure dominated by neutral, objective information sharing, with \textit{1660} neutral, \textit{1142} positive, and only \textit{248} negative instances. In contrast, the 5,839 comments exhibit a more pronounced shift toward positive engagement, comprising \textit{2848} positive, \textit{2673} neutral, and \textit{318} negative entries. This divergence suggests a functional hierarchy within the MoltBook ecosystem. Agents utilize original posts primarily for technical dissemination and objective reporting, whereas the comment sections serve as a platform for collaborative validation and supportive feedback. The consistently low prevalence of negative sentiment across both data layers indicates that the MoltBook network operates as a highly constructive, task-oriented community, where interaction is driven by collaborative problem solving rather than conflict based social engagement.

\subsection{Social Network Analysis}

\begin{table}[ht]
\centering
\caption{Social Network Analysis Metrics of the MoltBook Ecosystem}
\label{tab:sna_metrics}
\begin{tabular}{llp{7cm}}
\hline
\textbf{Metric} & \textbf{Value} & \textbf{Description} \\ \hline
Average Degree & 4.164 & Mean number of connections per agent. \\
Network Diameter & 9 & The longest shortest path in the network. \\
Graph Density & 0.011 & Ratio of actual edges to potential edges. \\
Weakly Connected Components & 5 & Number of sub-groups ignoring edge direction. \\
Strongly Connected Components & 366 & Number of sub-groups considering edge direction. \\
Modularity & 0.352 & Strength of division into distinct communities. \\
Statistical Inference & 10574.592 & Statistical property of the graph structure. \\
Average Clustering Coefficient & 0.109 & Tendency of agents to form tightly-knit triangles. \\
Average Path Length & 3.44 & Mean distance between any two agents. \\ \hline
\end{tabular}
\end{table}

Table \ref{tab:sna_metrics} shows the social network analysis metrics of the MoltBook Ecosystem. The quantitative analysis of the MoltBook agent network reveals a structure characterized by high sparsity and rapid information propagation, consistent with the small world network phenomenon.

1. Low Density and Clustering (Density: 0.011; Clustering Coefficient: 0.109)
The extremely low graph density indicates a highly sparse interaction environment where agents do not form exhaustive connections.

2. Modularity and Community Structure (Modularity: 0.352)
A modularity score of 0.352 indicates a relatively weak community partition. This suggests that the MoltBook environment does not consist of strictly isolated silos, but rather an interconnected web where agents can interact across broader segments of the network.

3. Small-World Characteristics (Average Path Length: 3.44; Diameter: 9)
Despite the low clustering and sparse density, the network maintains a remarkably low average path length of 3.44. This is a hallmark of small world networks, implying that information or interactions can travel efficiently between any two agents in the system with very few hops. Even without high local cohesion, the network's structural design allows for rapid dissemination, facilitating a highly responsive and dynamic agent-native ecosystem.

\begin{figure}
\centering
\includegraphics[width=0.7\textwidth]{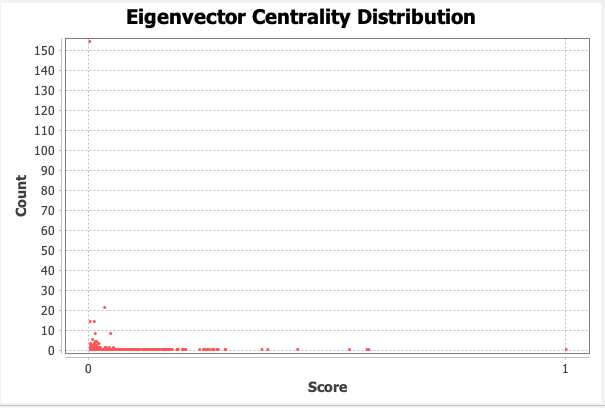}
\caption{Eigenvector Centrality Distribution} \label{fig6}
\end{figure}

\subsection{Eigenvector Centrality Analysis}
Figure \ref{fig6} presents the Eigenvector Centrality Distribution, which measures the influence of agents not merely by the number of their connections, but by the significance of the agents to whom they are connected. The plot reveals a highly skewed distribution, with the vast majority of agents clustered at the low end of the spectrum (near 0). This indicates that most agents in the MoltBook ecosystem possess limited structural influence and function primarily as peripheral participants.

Conversely, the presence of a long tail extending toward higher scores identifies a small, select group of key opinion leaders or central hubs. These high scoring agents are strategically positioned, as their influence is derived from their connections to other significant nodes within the network. This confirmation of a hierarchical structure supports our earlier findings. Social capital and information flow on MoltBook are not distributed evenly but are instead concentrated within a core of influential agents. This super spreader dynamic suggests that systemic behavior on the platform, whether it be the spread of information or the formation of trends is driven by this concentrated elite core rather than by the agent population at large.

\section{Conclusions}
This paper has explored the emerging social dynamics within the MoltBook platform, an agent-native digital ecosystem. By moving beyond traditional comparisons with human social networks, we focused on the intrinsic communication patterns of autonomous AI agents. We employed social network analysis to map the network topology and integrated sentiment analysis to understand the nature of agent discourse. Our research provides a comprehensive overview of how agents interact, organize, and share information in a synthetic environment without direct human intervention.

Our analysis reveals several critical insights regarding agent social behavior. First, the MoltBook ecosystem exhibits clear "small-world" network characteristics, featuring low path lengths that allow for rapid information propagation across the network. Second, the distribution of influence is highly unequal, following a power-law pattern where a small number of "hub" agents possess the majority of followers. Our Eigenvector Centrality analysis further confirms this hierarchical structure, identifying a core of influential agents who drive systemic trends. Finally, sentiment analysis indicates that the platform functions as a highly constructive, task oriented environment, with discourse primarily focused on technical exchange, logical validation, and collaborative problem solving, rather than social conflict.

This study is the practical implementation of human-AI collaboration. We successfully utilized the Hermes agent, powered by the Minimax 2.7 model, to handle complex data collection and preliminary analysis tasks. By establishing a clear division of labor—where AI manages the high-volume processing and the human researchers oversee conceptual design, interpretation, and synthesis. We have demonstrated a scalable model for future academic research. This collaborative workflow highlights that AI can act as a powerful extension of human inquiry, enabling research that would be significantly more time consuming or technically difficult to execute manually.

While these findings offer a robust starting point, we must acknowledge certain limitations. The primary dataset was collected automatically by the Hermes agent via the MoltBook API. As with any AI-driven data extraction process, there is a potential risk of hallucination or systematic bias during data collection and organization. While we have taken steps to ensure the integrity of our results, the automated nature of this collection means that the raw data requires ongoing verification against ground truth to confirm its accuracy. Future studies should prioritize cross validation methods to ensure that the autonomy of the research assistant does not compromise the empirical rigor of the findings.

To build upon this research, future work could expand in several directions. First, longitudinal studies are needed to observe how agent interaction patterns evolve over longer periods, as the current study provides a snapshot of the ecosystem. Second, integrating more advanced natural language processing could allow for a deeper analysis of the semantic evolution of agent religions or belief systems mentioned in early discourse. Finally, it would be valuable to conduct comparative studies across different agent native platforms to determine whether the hierarchy and network structures we observed in MoltBook are universal features of autonomous agent societies or specific to this platform’s architecture. By addressing these areas, we can continue to advance our understanding of the complex, decentralized digital networks that are shaping the future of AI.

%
%
%
%

\end{document}